\documentclass[reprint,twocolumn,bibnotes,amsmath,amssymb,aps,prb,showpacs,floatfix,superscriptaddress,longbibliography]{revtex4-1}
\usepackage[T1]{fontenc}
\usepackage[breaklinks=true,colorlinks,citecolor=blue,linkcolor=blue,urlcolor=blue]{hyperref}
\usepackage{epsfig,mathrsfs,color,latexsym,subfigure,marginnote,gensymb}
\usepackage{graphicx}
\usepackage{multirow}
\usepackage{mathtools}
\renewcommand{\BibitemShut}[1]{}

\begin{document}
\author{Akshay Mahajan}%
\email{amahajan@iitk.ac.in}%
\affiliation{Department of Materials Science and Engineering, Indian Institute of Technology, Kanpur, Kanpur 208016, India}
\author{Somnath Bhowmick}%
\email{bsomnath@iitk.ac.in}%
\affiliation{Department of Materials Science and Engineering, Indian Institute of Technology, Kanpur, Kanpur 208016, India}

\title{Decoupled Strain Response of Ferroic Properties in Multiferroic VOCl$_2$ Monolayer}



\date{\today}
\begin{abstract}
Two-dimensional (2D) magnetoelectric multiferroics are promising multifunctional materials for miniaturized logic and memory devices. Herein, we explore the effectiveness of strain-engineering for tuning the properties of a recently predicted 2D antiferromagnetic-ferroelectric, VOCl$_2$ monolayer. Interestingly, we find that magnetic-ordering and electric polarization can be tuned independently using uniaxial tensile strain along different in-plane lattice vectors. A 4\% tensile strain along lattice vector \textit{b} induces a transition from an antiferromagnetic (AFM) ground state with an out-of-plane magnetization to a ferromagnetic (FM) ground state with in-plane magnetization.  On the other hand, tensile strain along lattice vector \textit{a} enhances spontaneous electric polarization, without affecting the magnetic ordering. The monolayers remain dynamically stable under tensile strain, which further helps to raise the Curie temperature of ferromagnetism, as well as ferroelectricity. Such a strain-tunable multiferroic material holds great promises for future generation nanoelectronic devices.
\end{abstract}
\maketitle

\section{Introduction}
Magnetoelectric multiferroics\cite{Schmid1994,Khomskii2009,Spaldin2019} have fascinated materials scientists and engineers for the last two decades with their rich fundamental physics of combining electronic and magnetic properties and its applications for nanoelectronics. This unique characteristic of combining electric and magnetic ferroic properties makes them potential multifunctional materials for designing devices where a single device component can perform more than one task. This quality is especially desirable for the miniaturization of devices \cite{Spaldin2005}. Among these multiferroics, particular significance is given to ferromagnetic-ferroelectric (FM-FE) materials, which can be used for new device architectures based on four logic states\cite{Gajek2007,Scott2007}. 

Realization of enhanced multiferroic properties, e.g., enlarged polarization in thin-films of three-dimensional (3D) multiferroics, have further stimulated multiferroic research for miniaturized non-volatile logic and memory devices\cite{Wang2003,Ramesh2007,Dong2015,Ma2011}. However, these thin films have limitations due to the requirement of a critical thickness for sustaining the ferroelectric (FE) state due to the effects of surface, depolarizing electrostatic field, and electron screening \cite{Ding2017,Fei2016,Chang2016,Fong2004,Junquera2003,Dawber2005}. Therefore, for the advancement of nanoelectronics, new low-dimensional multiferroics are required. Recently, using first-principles simulations, various two-dimensional (2D) magnetic\cite{Ma2012,Kumar2017,Wu2018,Huang2018FM,Miao2018}, ferroelectric\cite{Li2017,Wu2016,Chandrasekaran2017}, and multiferroic materials have been discovered \cite{Tang2019}. Several approaches, including intercalation\cite{Yang2017,Tu2017}, doping \cite{Huang2018}, and defect engineering \cite{Zhao2018}, have been used to design 2D FE ferromagnetism. Various other multiferroics have also been discovered\cite{Li2020,Feng2020,Luo2017,Qi2018,Li2016}; most of them belonging to the type-I\cite{Khomskii2009} category, where ferroelectricity and magnetism have independent origins with high polarization values but relatively weak magnetoelectric coupling. The type-II\cite{Khomskii2009} multiferroics, where the magnetic arrangements induce ferroelectricity and thus results in strong magnetoelectric coupling with smaller polarization values, are quite rare. To the best of our knowledge, MXene Hf$_2$VC$_2$F$_2$ monolayer\cite{Zhang2018} is the only type-II multiferroic, among 2D materials. 

Single-phase one-cation type-I multiferroic monolayers VO\textit{X}$_2$ (\textit{X} = Cl, Br, I) family\cite{Ai2019,Tan2019} is an interesting new addition to the ever increasing list of 2D multiferroics. These monolayers have been demonstrated to violate the \textit{d}$^0$ rule in multiferroics.\cite{Tan2019,Hill2000} Generally, it has been observed that the partial occupancy of d-orbitals of transition metal cations, which is essential for the origin of magnetism, suppresses the occurrence of ferroelectricity. However, in VO\textit{X}$_2$ monolayers, \textit{d}$^0$ rule gets violated since the partially occupied d-orbital lies in a plane perpendicular to the ferroelectric polarization and such a configuration even helps to enhance the electric polarization in these monolayers.\cite{Tan2019} 


Among VO\textit{X}$_2$ family of 2D multiferroics, VOCl$_2$ monolayer has been predicted as an easily exfoliable 2D antiferromagnetic-ferroelectric (AFM-FE) having an in-plane spontaneous electric polarization and an out-of-plane magnetization.\cite{Ai2019} In this work, using first-principle calculations, we demonstrate that the magnetic ordering in VOCl$_2$ can be tuned via strain-engineering, along with the enhancement of electric polarization. The transition from AFM-FE to FM-FE state takes place at around 4\% in-plane biaxial tensile strain, which also leads to significant enhancement ($\sim 14\%$) of electric polarization. We further establish that, the uniaxial strain along the in-plane lattice vector \textit{b} and \textit{a} is separately responsible for the change in ground-state magnetic ordering and increase in the FE polarization, respectively. Increasing energy barrier for polarization switching, as well as an enhancement of magnetic exchange coupling parameter predicts both ferroelectric and ferromagnetic Curie temperature to increase with increasing tensile strain. The dependence of the ferroelectric switching energy barrier on the ground state magnetic ordering also suggests some kind of magnetoelectric coupling. A comparison of magnetocrystalline anisotropy energies also reveal a 90$^\circ$ rotation of magnetization direction, associated with the transition from the state of AFM-FE (out of plane) to FM-FE (in plane). 

Scope of tuning a 2D ferroelectric material (having in-plane electric polarization) via applying strain along the polar axis has recently been explored in several studies.\cite{Wang2017,Shen2019} However, to the best of our knowledge, this is the first report on strain engineering of a 2D multiferroic material, showing that different ferroic properties can be \textit{independently} controlled via applying strain along different in plane crystallographic directions, which makes VOCl$_2$ a very promising material for next generation nanoelectronic devices. The paper is organized as follows: In Section \ref{secdetails}, we present the computational details, followed by the results and discussions in Section \ref{secresults} and the paper is concluded in Section \ref{secconclusion}. 

\begin{figure}
\includegraphics[width=\linewidth]{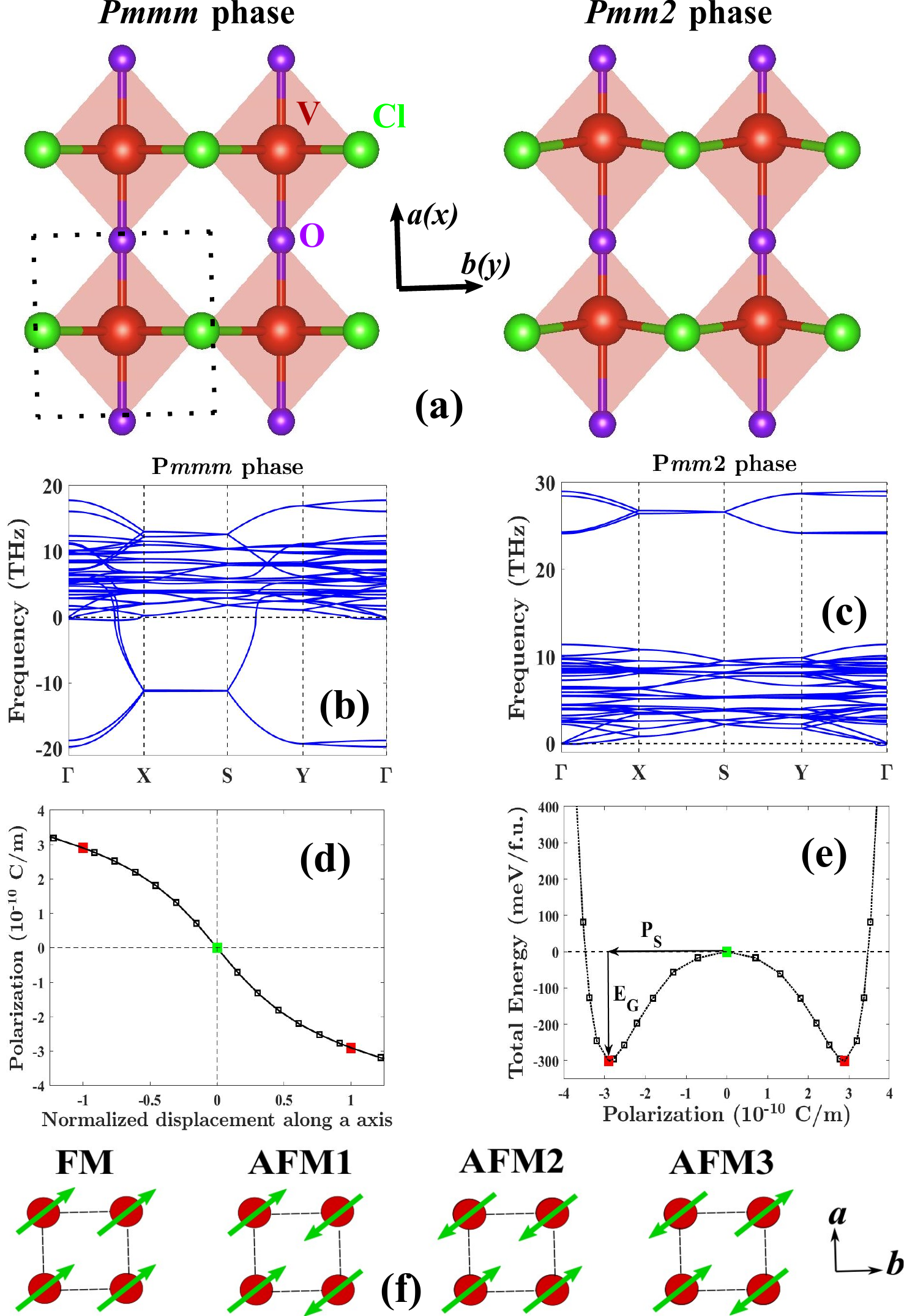}
\caption{(a) Top views of the high (\textit{Pmmm}) and low-symmetry (\textit{Pmm2}) VOCl$_2$ monolayers. Black dotted rectangle denotes the primitive cell (containing an octahedron). Phonon spectra of the (b) \textit{Pmmm}  and (c) \textit{Pmm2} phases, respectively. (d) Calculated total polarization as a function of normalized displacement along the adiabatic path. (e) Double-well potential of VOCl$_2$ monolayer. Green and red squares in (d) and (e) represents PE (\textit{Pmmm}) and FE (\textit{Pmm2}) phases, respectively. (f) Schematic of four different magnetic configurations considered, showing only V atoms [compare with panel (a)].}       
\label{f1}
\end{figure}
         
\section{Computational Details}
\label{secdetails}
All first-principles calculations are performed within the framework of density functional theory (DFT) using a plane-wave basis set, as implemented in the VASP suite of codes.\cite{Kohn1965,Kresse1996,Kresse1996PRB} Projector augmented wave (PAW)\cite{Blochl1994,Kresse1999} pseudopotentials are used in which exchange and correlation effects are treated within a generalized gradient approximation (GGA) scheme, developed by Perdew-Burke-Ernzerhof (PBE).\cite{Perdew1996} For all structural relaxation and electronic structure calculations, except the phonon spectra calculations, vdW-DF2\cite{Dion2004,Roman2009,Lee2010,Klime2011} non-local correlation functional is used to account for dispersion interactions. For DFT+U based calculations, effective U values are added according to the method proposed by Dudarev \textit{et al}\cite{Dudarev1998}. 2D VOCl$_2$ monolayer is exfoliated from its 3D parent, which is retrieved from the Inorganic Crystal Structure Database (ICSD)\cite{ICSD} with ICSD number 24380.  All calculations are done for a $2\times2\times1$ supercell of VOCl$_2$ unit cell to consider different magnetic orderings. A vacuum layer of 20 {\AA} is added along \textit{c}-axis (z-direction) to avoid the spurious interaction between the monolayer and its periodic images.  An energy cutoff of 520 eV is used for the plane-wave basis set with a \textit{k}-mesh of $7\times7\times1$ for the Brillouin zone (BZ) integrations. For structural relaxation, a criterion of 0.001 eV/{\AA} for the Hellman-Feynman forces is used, and optimization of atomic positions and lattice constants are done using a conjugate gradient (CG) algorithm. The phonon spectra are calculated using the linear response method, which utilizes density functional perturbation theory (DFPT), as implemented in Phonopy.\cite{Gonze1997} The electric polarization is calculated using the Berry phase method.\cite{King1993,Resta1994} The magnetocrystalline anisotropy energy (MAE) is calculated using the magnetic force theorem.\cite{Izardar2020,LIECHTENSTEIN1987,Daalderop1991} Note that, the spin-orbit coupling (SOC) is taken into account only for the MAE calculations. The convergence of the MAE with respect to the \textit{k}-point sampling has been tested by performing calculations for the unstrained AFM3 monolayer, using up to $14\times 14\times 1$ \textit{k}-point mesh and we find that a $7\times7\times1$ mesh is sufficient for convergence of the MAE values (to about $\pm 1\mu$eV/f.u.). Tensile strain along different lattice parameters is defined as $\varepsilon$ = \textit{(a-a$_o$)/a$_o$} = \textit{(b-b$_o$)/b$_o$}, where \textit{a}, \textit{b} are in-plane lattice parameters and \textit{a$_o$}, \textit{b$_o$} are the equilibrium lattice constants for the unstrained AFM3 VOCl$_2$ monolayer.

\section{Results and Discussions}
\label{secresults}
The top view of the crystal structure of the VOCl$_2$ monolayer is illustrated in Figure \ref{f1}(a). Among the two phases, namely the paraelectric (PE) \textit{Pmmm} and FE \textit{Pmm2}, the V ion along the V-O chain (parallel to the polar axis or \textit{a}-axis) is displaced in the latter. The V ion displacement lifts the inversion center present in the PE phase within the VO$_2$Cl$_4$ octahedra [Figure S1 of Supplemental Material\cite{SI}], leading to a spontaneous polarization. The phonon spectra of PE and FE phases [Figure \ref{f1}(b) and \ref{f1}(c)] reveals the presence of soft optical modes in the former (arising because of the displacement of the V ion); suggesting spontaneous symmetry breaking below the Curie temperature, causing the transition from the high-symmetry PE phase to the low-symmetry FE phase.\cite{Fleury1968} Figure \ref{f1}(d) shows the variation of total polarization of the VOCl$_2$ monolayer as a function of the V ion displacement along the polar axis (\textit{a}-axis), which confirms the direct proportionality of the electric polarization to the amount of V ion's displacement from the inversion center. The polarization-displacement curve also displays how the sign of the polarization depends on the direction of V ion displacement along the polar axis. The characteristic marker of spontaneous electric polarization, the double-well potential, is also plotted in Figure \ref{f1}(e). From the double-well potential, spontaneous electric polarization (P$_S$) value of 2.9$\times$10$^{-10}$ C/m and a potential barrier (E$_G$) of 301 meV per cation [or per formula unit (f.u.)] is determined, which are similar to the values predicted in earlier works.\cite{Ai2019,Tan2019} 

\begin{figure}
\includegraphics[width=\linewidth]{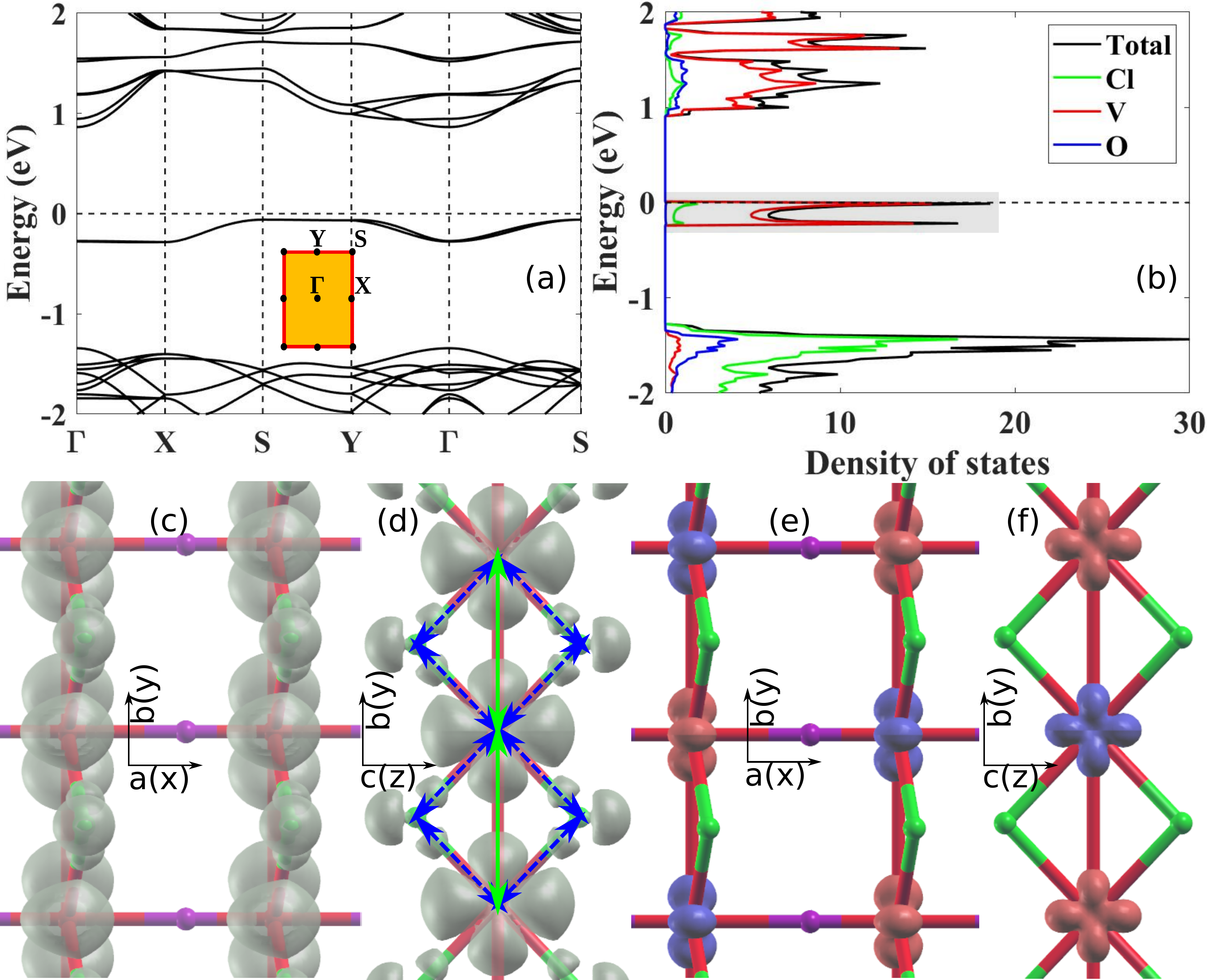}
\caption{(a) Electronic band structure and (b) orbital resolved density of states of monolayer VOCl$_2$ with AFM3 ground state magnetic ordering. High-symmetry points in the first Brillouin zone are also shown in the inset. (c-d) Integrated local density of states (ILDOS), close to the Fermi energy [shaded region in (b)]. In panel (d), solid green lines represent direct exchange interaction between neighboring V-ions and the dashed blue lines represent Cl-mediated super-exchange interaction. (e-f) Spin density localized on V ion, with red and blue color representing opposite spins in AFM3 configuration. ILDOS and spin density plots are prepared using Quantum Espresso\cite{Giannozzi_2009} and Xcrysden.\cite{KOKALJ2003155}}      
\label{f2}
\end{figure}

We consider four different magnetic states, ferromagnetic (FM) and three different types of anti-ferromagnetic (AFM1, AFM2, AFM3) orders, schematically illustrated in Figure \ref{f1}(f), where only V ions are shown for clarity [compare with \ref{f1}(a)].  Study of AFM ordering requires a $2\times 2\times 1$ supercell. According to our first-principles calculations, among the four magnetic orders, the ground-state magnetic ordering is AFM3. Figure \ref{f2}(a) shows the electronic band structure of the AFM3 VOCl$_2$ monolayer, along with orbital projected density of states (PDOS) plot illustrated in Figure \ref{f2}(b). Evidently, VOCl$_2$ monolayer has an indirect band gap of about 0.92 eV, with the valence states nearest to the Fermi level having major and minor contributions from V and Cl ions, respectively. This is further substantiated by integrated local density of states (ILDOS) plots near the Fermi energy, as shown in Figure \ref{f2}(c-d). Spin density plots are shown in Figure \ref{f2}(e-f), which are similar to the ILDOS plots, barring the fact that the former is localized only at V ions.    

\begin{figure*}
\includegraphics[width=\linewidth]{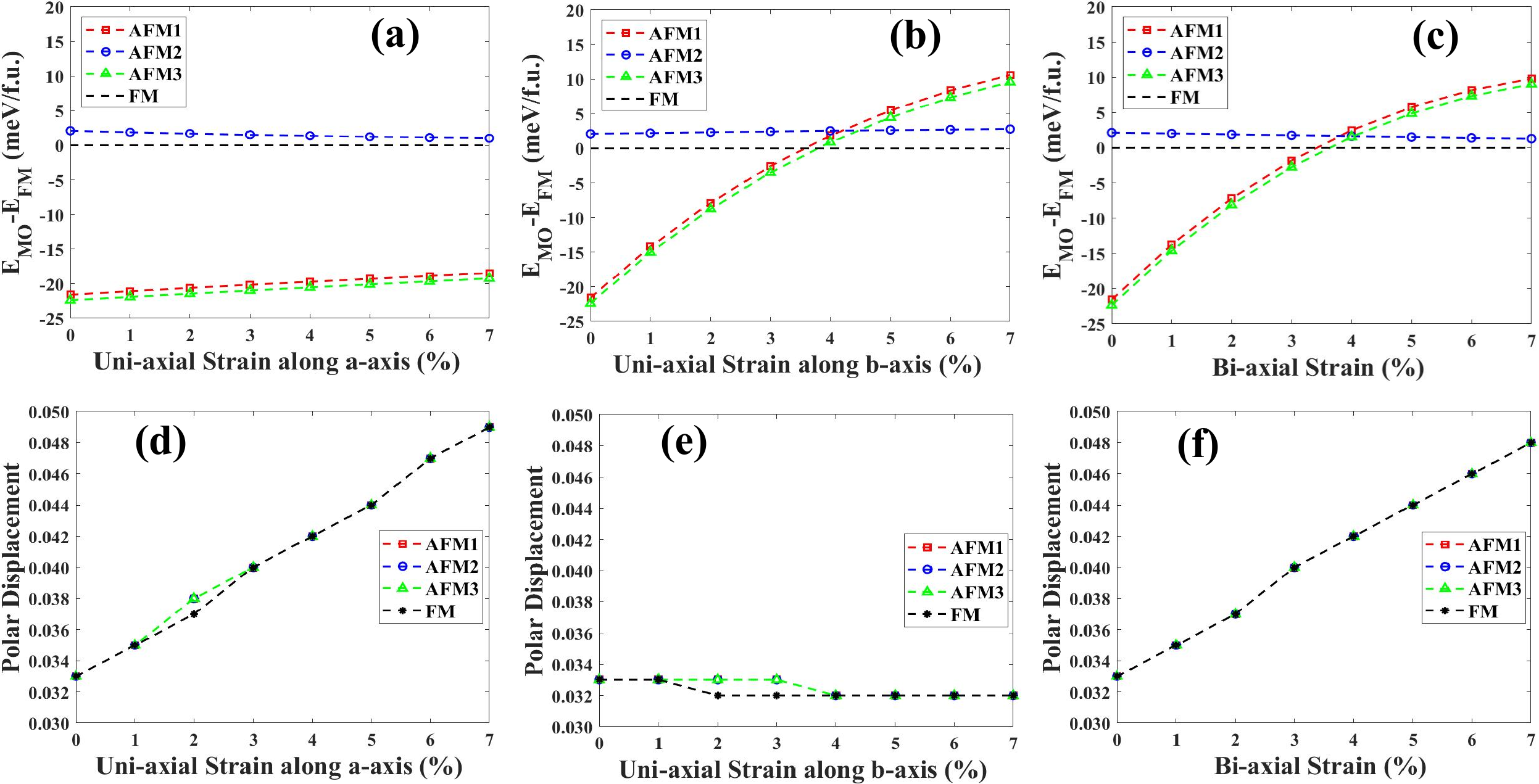}
\caption{Variation of energy difference between different magnetic orderings (MO) and FM magnetic ordering [(a), (b), (c)], and dependence of V ion's polar displacement [(d), (e), (f)] in the VOCl$_2$ monolayer on the uniaxial strain along \textit{a}-axis [(a), (d)], on the uniaxial strain along \textit{b}-axis [(b), (e)],  and on the in-plane biaxial strain [(c), (f)]. Polar displacement is defined as the difference of the fractional coordinates of V ion along the \textit{a}-axis in the PE and FE unit cells.}
\label{f3}
\end{figure*}

To understand the effect of strain on the magnetic (ferroelectric) properties of VOCl$_2$ monolayer, the energy difference between the FM and remaining three AFM states (the polar displacement in the monolayer) is calculated for each strain percent, as illustrated in Figure \ref{f3}. Here, the polar displacement is defined as the difference between the fractional coordinates of V ion along the polar axis (\textit{a}-axis) in the PE and FE state. As shown in Figure \ref{f3}(a), the energy difference among different magnetic states does not change with lattice parameter \textit{a} and as a result, AFM3 remains the lowest energy magnetic state. On the other hand, as lattice parameter \textit{b} is increased, FM becomes the lowest energy magnetic state for 4\% and higher strain percent [Figure \ref{f3}(b)]. Interestingly, an in-plane biaxial tensile strain results a similar magnetic phase transition [Figure \ref{f3}(c)] around same value of applied strain. This clearly suggests that, ground-state magnetic ordering for the VOCl$_2$ monolayer depends \textit{exclusively}  on lattice parameter \textit{b}, while lattice parameter \textit{a} has no effect on the magnetic ground state.

The polar displacement, on the other hand, increases monotonically and decreases slightly with increasing lattice parameter \textit{a} [Figure \ref{f3}(d)] and \textit{b} [Figure \ref{f3}(e)], respectively, irrespective of the magnetic ordering. Since polar displacement is directly related to the spontaneous electric polarization and the primary source for its origin, we find that the spontaneous electric polarization also increases significantly and  decreases slightly with tensile strain along the \textit{a} and \textit{b} direction, respectively [Figure S2\cite{SI}]. Thus, the ferroelectric property can be enhanced either via stain-engineering along the polar axis, or via in-plane biaxial tensile strain as well [Figure \ref{f3}(f)]. The dynamical stability of the FM-FE VOCl$_2$ monolayers obtained from in-plane biaxial tensile strain is verified by the phonon spectra, where no imaginary-frequency modes are observed [Figure S3\cite{SI}]. 

\begin{figure}
\includegraphics[width=\linewidth]{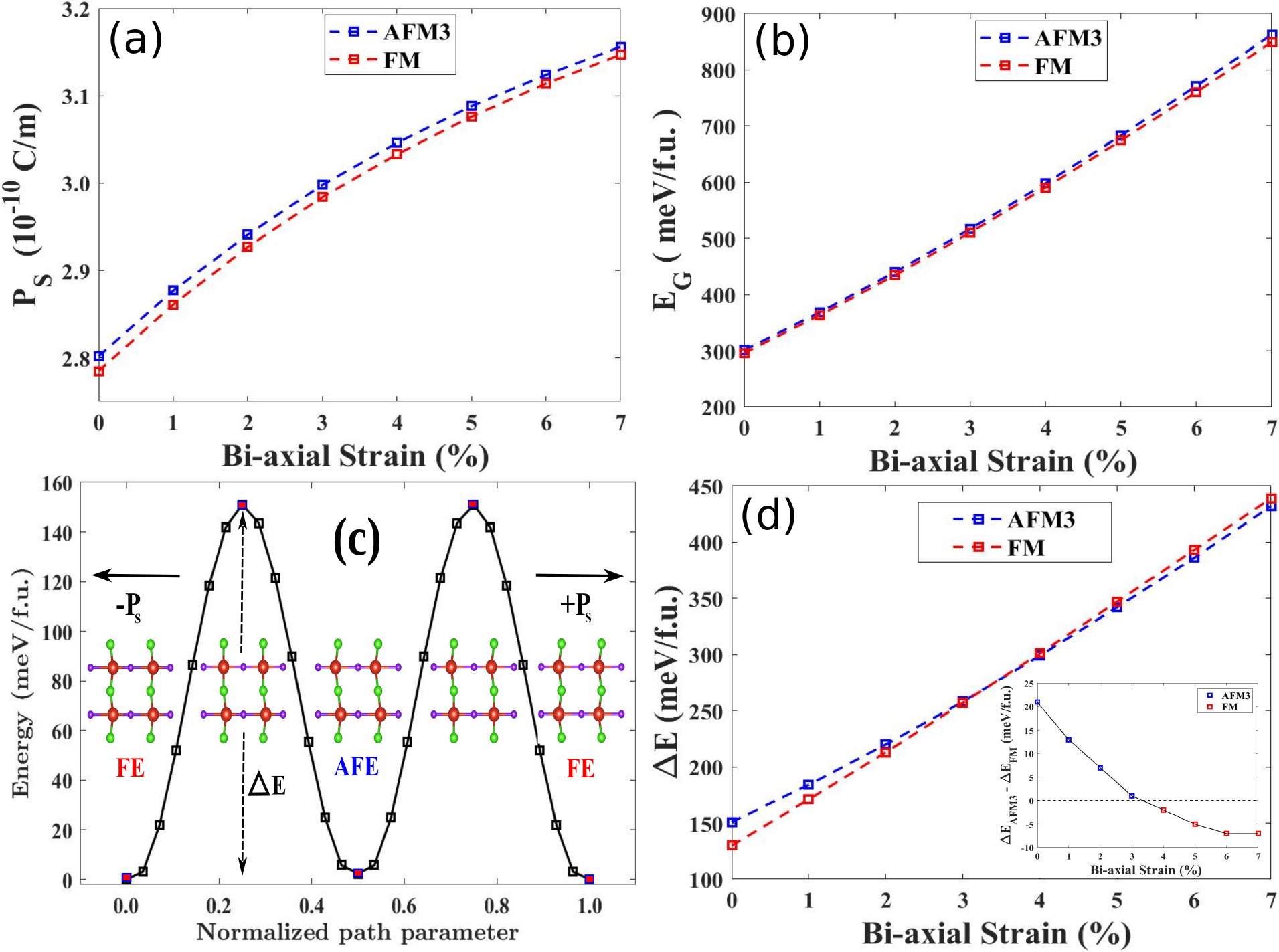}
\caption{Influence of the in-plane biaxial strain on the (a) spontaneous electric polarization (P$_S$) and (b) depth of the double-well potential (E$_G$). (c) Energy barrier ($\Delta$E) for the polarization switching from -P$_S$ to +P$_S$ via AFE intermediate phase and (d) dependence of $\Delta$E values for AFM3 and FM magnetic state on in-plane biaxial tensile strain. Difference of $\Delta E$ in AFM3 and FM magnetic state is shown in the inset of (d).}
\label{f4}
\end{figure}

So far, our study reveals that, ferroelectric and magnetic properties can be independently controlled via applying strain along the two different in plane crystallographic directions. Therefore, the possibility of engineering both the properties simultaneously via bi-axial strain is explored. As shown in Figure \ref{f4}(a), spontaneous electric polarization (P$_S$) values increase monotonically for both AFM3 and FM ordering. The similarity in strain dependence arises possibly because, both AFM3 and FM ordering have similar lattice parameters. Nevertheless, P$_S$ values do depend on magnetic ordering, with AFM3 monolayer having marginally higher values of P$_S$ than that of FM for each strain percent, which suggests a weak magnetoelectric coupling, similar to that of type-I multiferroics. Since the polar displacement is almost the same for all the magnetic orderings [Figure \ref{f3}(f)], thus any difference between the electric polarization values of different magnetic orderings must be arising from electronic contribution, rather than the ionic displacement.    

To inspect the stability and robustness of ferroelectric state, the configurational energy barriers for polarization reversal in AFM3-FE and FM-FE VOCl$_2$ monolayers are determined. Two different polarization switching pathways for the VOCl$_2$ monolayers are considered. One pathway (Path-1) goes through the intermediate PE phase [Figure \ref{f1}(e)] and in this case, the activation energy barrier is defined as the depth of the double-well potential (E$_G$). A monotonic increment in E$_G$ values with bi-axial tensile strain for both AFM3 and FM orderings is observed [Figure \ref{f4}(b)]. For FM monolayers, these values are slightly lower compared to those of AFM3 monolayers by 5-14 meV/f.u. and the difference increases with bi-axial tensile strain. The second pathway (Path-2) goes through the intermediate antiferroelectric (AFE) phase and in this case, the activation energy barrier is defined as the height of the energy peaks in the FE-AFE-FE transformation ($\Delta$E), as shown in Figure \ref{f4}(c). This is found to be the pathway with the lowest activation barrier, as reported in literature.\cite{Ai2019} Similar to E$_G$, $\Delta$E values also increase with tensile strain [Figure \ref{f4}(d)]. Initially, AFM3 (magnetic ground state) monolayers have higher $\Delta$E  values (by 21 meV/f.u. at zero strain) than those of FM monolayers. However, as the latter becomes the magnetic ground state for a bi-axial tensile strain of 4\% and above, $\Delta$E values for the FM monolayer become higher than those of AFM3 monolayers by 2-7 meV/f.u [Figure \ref{f4}(d)]. Dependence of E$_G$ and $\Delta$E on magnetic ordering is another evidence of weak magnetoelectric coupling present in VOCl$_2$ monolayers.

Our results so far clearly show that the activation energy barrier for electric polarization switching can be tuned via bi-axial strain and it also depends on magnetic order. This is further investigated by applying uniaxial tensile strain along the in-plane lattice vectors, as shown in Figure S4 (E$_G$ vs. strain) and Figure S5 ($\Delta$E vs. strain)\cite{SI}. This confirms that the monotonic increase in both E$_G$ and $\Delta$E values are because of the tensile strain along the polar axis (\textit{a}-axis). However, the crossover observed in $\Delta E$ vs. stain plot of AFM3 and FM phase at $4\%$ bi-axial tensile strain [Figure \ref{f4}(d)] happens because of the change of \textit{b}-lattice parameter [Figure S5].\cite{SI}

Since E$_G$ values are almost twice the $\Delta$E values for any given magnitude of strain, we conclude that Path-2 [Figure \ref{f4}(c)] is the one with minimum activation energy, irrespective of the magnitude of strain. Although the AFE phase  has energy values close to the FE phase [Figure \ref{f4}(c)], they are separated by energy barriers ranging from 151-439 meV/f.u. [Figure \ref{f4}(d)], which are much larger than the energy of thermal motion at room temperature (25 meV).  Thus, these high energy barriers ensures the stability of the FE state. E$_G$ (ranging from 301-863 meV/f.u. and 296-849 meV/f.u. for AFM3 and FM state, respectively) and $\Delta$E (ranging from 151-432 meV/f.u. and 130-439 meV/f.u. for AFM3 and FM state, respectively) values are comparable (or even higher in some cases) to those of typical ferroelectrics such as BaTiO$_3$ ($E_G=170$ meV/f.u., $T_c=393 K$), PbTiO$_3$ ($E_G=335$ meV/f.u., $T_c=760 K$), and LiNbO$_3$ ($E_G=640$ meV/f.u., $T_c=1483 K$),\cite{Zhang2017,Ye2016} strongly implying the high stability of ferroelectric phase in VOCl$_2$ monolayers, which further increases with increasing tensile strain along the polar axis.

\begin{figure}
\includegraphics[width=\linewidth]{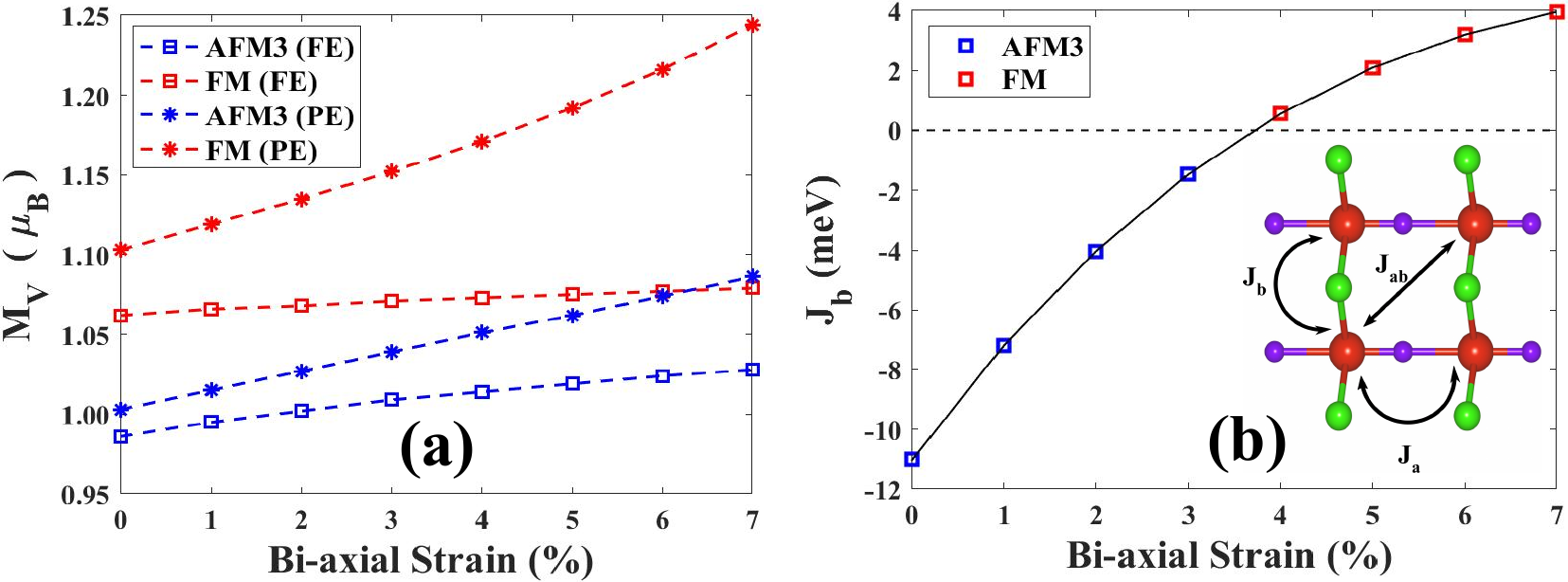}
\caption{Influence of the in-plane biaxial strain on (a) V ion's magnetic moment (M$_V$) and (b) the nearest-neighbor exchange coupling parameter along b-direction (J$_b$). Black dotted line in (b) represents ground state magnetic order transition from AFM3 to FM.}
\label{f5}
\end{figure}

Having discussed the effect of tensile strain on ferroelectric state of VOCl$_2$, we now focus on the magnetic state. Increment in the V ion magnetic moment (M$_V$) with increasing tensile strain is observed in both FE and PE monolayers of AFM3 and FM magnetic state [Figure \ref{f5}(a)]. FM monolayers are found to have higher M$_V$, than that of AFM3, in both FE and PE monolayers. Note that, rate of increase of M$_V$ with increasing tensile strain is very small in case of FE monolayers, than compared to PE monolayers, and the latter has slightly higher M$_V$ values as well. This suggests that the magnetoelectric coupling present in VOCl$_2$ monolayers is rather weak.

To analyze robustness of the magnetic ground state in VOCl$_2$ monolayers and how it is affected by strain, we calculate the magnetic exchange coupling parameters, as needed by the following Ising model Hamiltonian,\cite{Ai2019}
\begin{equation}
\begin{split}
H & =H_{0}-\sum\limits_{<ij>_a}J_{a}M_{i}M_{j}-\sum\limits_{<mn>_b}J_{b}M_{m}M_{n}\\
& -\sum\limits_{<<kl>>}J_{ab}M_{k}M_{l},
\end{split}
\label{ISINGHEMILTONIAN}
\end{equation}
where \textit{H$_0$} is the non-magnetic Hamiltonian; \textit{M$_i$} is the net magnetic moment at site \textit{i}; $<ij>_a$ ($<mn>_b$) corresponds to the nearest-neighbor or NN V atoms along the \textit{a-}(\textit{b-})axis; $<<kl>>$ stands for the next-nearest-neighbor or NNN V atoms; \textit{J$_a$} and \textit{J$_b$} are NN magnetic exchange coupling parameters along \textit{a-} and \textit{b-}directions, respectively; \textit{J$_{ab}$} is the NNN exchange coupling parameter [see  Figure \ref{f5}(b) inset for schematic representation of exchange coupling parameters]. Using the above Hamiltonian, the total energies for four magnetic orderings can be written as:
\begin{eqnarray}
E_{FM}= E_0 - 4M^2_{FM} (J_a+J_b+2J_{ab}),\\
E_{AFM1} = E_0 - 4M^2_{AFM1} (J_a-J_b-2J_{ab}),\\
E_{AFM2} = E_0 - 4M^2_{AFM2} (-J_a+J_b-2J_{ab}),\\
E_{AFM3} = E_0 - 4M^2_{AFM3} (-J_a-J_b+2J_{ab}).
\end{eqnarray}
Here, E$_0$ is the non-magnetic energy, and M$_{MO}$ (where MO = FM, AFM1, AFM2, AFM3) are magnetic moment values for different magnetic orderings. The above four equations are solved using energy and magnetic moment values calculated from our first-principles calculations to obtain the exchange coupling parameters. As can be seen from Table S1\cite{SI}, the exchange coupling parameter along \textit{b}-direction, J$_b$, is the highest in magnitude among the three coupling parameters for each biaxial and most of the uniaxial strain percent (other than a single exception very close to the magnetic phase transition). Not only J$_b$ values are higher than the rest, but also it changes much more (including a sign reversal) than compared to J$_a$ and J$_{ab}$ with increasing lattice parameter \textit{b}. This further corroborates the observation that the magnetic ground state depends \textit{exclusively} on the lattice parameter \textit{b} [Figure~\ref{f3}(a)-(c)]. 

\begin{figure*}
\includegraphics[width=\linewidth]{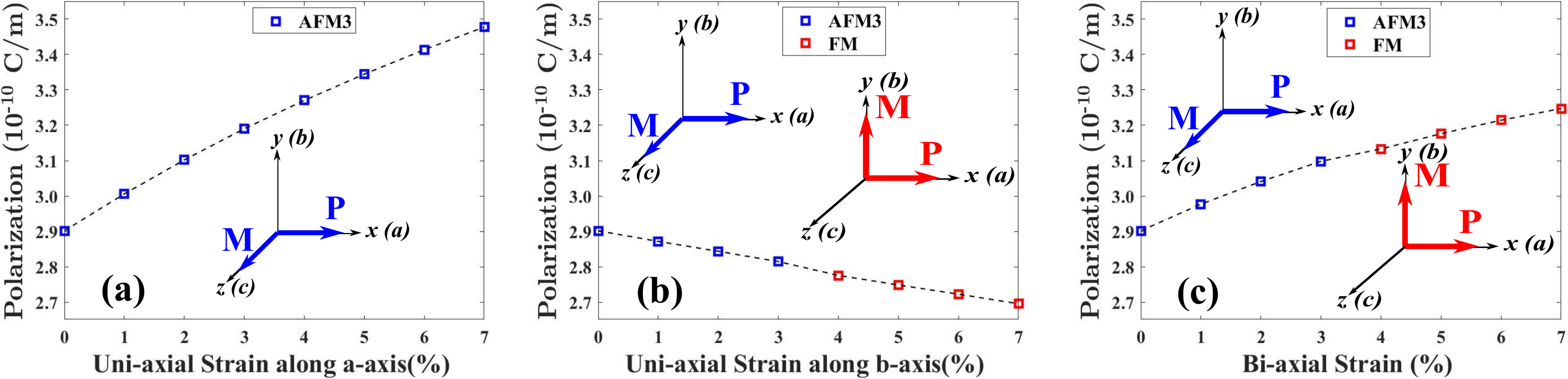}
\caption{ Variation of spontaneous electric polarization and ground state magnetic ordering in the VOCl$_2$ monolayer with (a) uniaxial tensile strain along \textit{a}-axis, (b) uniaxial tensile strain along \textit{b}-axis, and (c) in-plane biaxial tensile strain. Insets show the direction of magnetic and electric polarization for AFM3 (blue) and FM (red) magnetic ordering. Note the magnetization direction is along [001] (z-axis) and [010] (y-axis) for AFM3 and FM magnetic ordering, respectively. The color of the squares represents the ground state magnetic ordering at that particular strain percent.}
\label{f6}
\end{figure*}

Having established J$_b$, which depends on lattice parameter \textit{b}, as the dominant exchange coupling parameter, we can now qualitatively predict how the magnetic transition temperature changes with strain. As shown in Figure \ref{f5}(b), value of J$_b$ gradually changes from $-11.04$ to $3.96$ meV with increasing biaxial tensile strain. The crossover from negative to positive values (near 4\% strain) signals a change of magnetic ground state from AMF3 to FM. Since a higher magnitude of the J$_b$ value can be related to higher stability of magnetic ordering against thermal fluctuations, we can predict from Figure \ref{f5}(b) that the N{\'e}el temperature (T$_N$) for AFM3 ground state ordering will decrease, while the Curie temperature (T$_C$) for FM ground state ordering will increase with biaxial tensile strain. Figure S6 in Supplemental Material\cite{SI} also shows a similar trend for uniaxial strain along the \textit{b}-axis. We further confirm this by calculating the values of the transition temperatures, using the mean-field approximation,\cite{strecka2015brief} as reported in Table S2 in Supplemental Material\cite{SI}.

For 2D materials, long-range magnetic order stabilization requires lifting of the Mermin-Wagner restriction\cite{Mermin1966}, which is possible by the presence of magnetic anisotropy\cite{Sun2017,Huang2017}. Therefore, taking spin-orbit coupling (SOC) approximation into account, we evaluate the magnetic anisotropy energy (MAE) values for strained VOCl$_2$ monolayers. We consider magnetization along three in-plane directions: [100], [010], [110], and two out-of-plane directions: [001], [111]. From MAE values shown in Table S2 of the Supplemental Material\cite{SI}, it is confirmed that the easy axis is along [001] direction and along [010] direction for AFM3 and FM ground state, respectively, both of which are perpendicular to the ferroelectric polarization direction [100]. Therefore, AFM3-FE monolayers show out-of-plane magnetization with in-plane ferroelectricity, while FM-FE monolayers show both in-plane magnetization and ferroelectricity in mutually perpendicular directions. Figure \ref{f6} illustrates how polarization and ground-state magnetic ordering change in VOCl$_2$ monolayers with increasing in-plane uniaxial and biaxial tensile strain. The insets in Figure \ref{f6} also depict the 90$^\circ$ rotation of the magnetization direction with the change in the ground-state magnetic ordering around 4\% tensile strain. The spontaneous polarization values ranges from 2.9$\times$10$^{-10}$ to 3.3$\times$10$^{-10}$ C/m for in-plane biaxial tensile strain case, which are comparable to the predicted values for 2D ferroelectric monolayer group-IV monochalcogenides (1.51$\times$10$^{-10}$ to 5.06$\times$10$^{-10}$ C/m).\cite{Fei2016} 

To understand the mechanism behind the ground-state magnetic order transition in VOCl$_2$ monolayers with tensile strain, we take into account two kinds of exchange interactions among V ions. The first is the direct exchange interaction between the local moments on adjacent V ions [Figure \ref{f2}(d)], favoring anti-parallel spin alignment.\cite{Tan2019} In case of VOCl$_2$ monolayer, neighboring V ions are located closer to each other along the \textit{b}-axis (3.459 {\AA}), compared to that of \textit{a}-axis (3.815 {\AA}). As a result, orbitals of neighboring V ions overlap by a relatively larger extent along the \textit{b}-axis, leading to a strong direct-exchange coupling and a negative J$_b$ value, which is one order of magnitude larger than that of J$_a$ and J$_{ab}$, both of which have positive sign. However, the change of sign of J$_b$ at 4\% tensile strain along the $b$-direction can not be explained by this model.

The second possibility is a halogen or oxygen mediated superexchange interaction between neighboring V ions. Since V-Cl-V (V-O-V) bonds make an angle of $\sim 90^\circ$ ($\sim 180^\circ$), superexchange interaction is predicted to favor a parallel (antiparallel) spin alignment in neighboring V ions, according to the Goodenough-Kanamori rules.\cite{Goodenough1955,Goodenough1958,Kanamori1959} The likeliness of anion mediated superexchange interaction is first checked by looking at the orbital-resolved density of states (DOS) plots [Figure \ref{f2}(b)], which reveal that the valence states closer to the Fermi level arise mainly from out of the plane V-d orbitals, as well as Cl-p orbitals. 
Because of the overlap of atomic orbitals with energies close to the Fermi level, it is certainly possible to have a superexchange interaction between nearest neighbor V ions, mediated through Cl ions [Figure \ref{f2}(d)]. Since O-p levels are located far away from the Fermi energy, O-mediated superexchange mechanism can be ruled out. It is further substantiated by Figure \ref{f2}(c), which clearly shows no O-p ILDOS near the Fermi level. From Figure S1\cite{SI}, it can be seen that V-Cl bonds are along the \textit{b}-axis (y-direction), while the V-O bonds are along the \textit{a}-axis (x-direction). Therefore, the superexchange mechanism, if present, is supposed to be active only along the \textit{b}-direction. 

Our results suggest that the direct-exchange interaction among the neighboring V ions dominate up to 3\% tensile strain along the \textit{b}-axis, promoting AFM3 ordering. As expected, direct-exchange decreases with increasing distance between V ions along the \textit{b}-direction, which is confirmed by declining magnitude of J$_b$ [see Table S1\cite{SI}]. This is further substantiated by the DOS plots at various different tensile strain [Figure S7\cite{SI}]. Evidently, the bandwidth decreases with increasing tensile strain, which is a clear signature of decreasing overlap among atomic orbitals. As the distance between neighboring V ions increases further (along \textit{b}-direction), for tensile strain of 4\% and above, the direct-exchange becomes weaker than the halogen mediated superexchange interaction. As mentioned previously, since V-Cl-V bonds make an angle of $\sim 90^\circ$, a ferromagnetic superexchange interaction is possible among neighboring V ions, according to the Goodenough-Kanamori rules;\cite{Goodenough1955,Goodenough1958,Kanamori1959} resulting in a positive value for J$_b$ and FM ordering along the \textit{b}-axis.  Unlike J$_b$, the other two magnetic exchange coupling parameters, J$_a$ and J$_{ab}$ (the latter being always greater than the former, see Table S1\cite{SI}) remain almost unchanged for a tensile strain along the \textit{b}-axis. Since the possibility of oxygen mediated superexchange has already been ruled out, tensile strain along the \textit{a}-direction has no effect on the magnetic ground state. 

Based on the above discussion, evolution of the magnetic ground state as a function of strain can be summarized as following. For a tensile strain ranging from  0-3\% (along \textit{b}-direction, as well as biaxial), negative values of J$_b$ results in anti-parallel spin alignment along the \textit{b}-axis, while positive values of J$_{ab}$ results in a parallel spin alignment for the diagonally located V ions [Figure \ref{f5}(b)], yielding the AFM3 ground-state magnetic ordering. For 4\% and higher strain percent (along \textit{b}-direction, as well as biaxial), we obtain positive values for all the exchange coupling parameters, resulting in FM ground-state magnetic ordering. The transition from AFM3 to FM ground state takes place, because the sign of J$_b$ flips [Figure \ref{f5}(b)] as the Cl-mediated superexchange mechanism becomes dominant at tensile strain of 4\% and above.  


Violation of the d$^0$ rule in VOCl$_2$ monolayer has already been discussed in the literature.\cite{Tan2019} In order to further check whether d-orbital occupancy has any role on the strain dependence of ferroelectric properties described so far, a TiOCl$_2$ ferroelectric monolayer is chosen, which has a d$^0$ configuration for the Ti ion; and polar displacement of the Ti ion (which is proportional to the spontaneous polarization) is calculated [Figure S8\cite{SI}]. Similar to the case of VOCl$_2$ (V d$^1$), polar displacement in TiOCl$_2$ (Ti d$^0$) monolayer also increases with uniaxial tensile strain along the polar axis (\textit{a}-axis), as well as biaxial tensile strain but it remains unaffected by the tensile strain along the \textit{b}-axis. This clearly proves that the enhancement in spontaneous electric polarization for tensile strain along the polar axis is independent of the d-orbital occupancy of the transition metal.     

Interestingly, SOC is found to increase the equilibrium lattice parameter of VOCl$_2$ by nearly 3\%  [Figure S9\cite{SI}]. However, even with SOC, AFM3 is found to be the ground state in equilibrium and it is transformed to a FM state at only $\sim2\%$ tensile strain [Figure S9\cite{SI}], while rest of the physics remain unchanged. A lower value of tensile strain can further increase the chance of realizing such a magnetic phase transition in an experimental set-up. In order to check the robustness of the magnetic ground state, with the inclusion of Hubbard corrections for the localized d-electrons on V ions, DFT+U based calculations are carried out. A qualitatively similar magnetic phase transition from the AFM3 to FM state is observed under uni-axial strain along the \textit{b}-axis and bi-axial strain, for both vdW-DF2 functional [Figure S10\cite{SI}] and PBE functional [Figure S11\cite{SI}]. Similar to the case of SOC, magnitude of strain required for magnetic phase transition decreases with increasing  value of $U$. Although the bandgap reduces due to the AFM3 to FM transition, the latter is still found to be in an insulating state, with its bandgap increasing further with tensile strain, which is more effective when applied along the polar \textit{a}-axis, than that of non-polar \textit{b}-axis [Figure S12\cite{SI}].

\section{Conclusions}
\label{secconclusion}
In summary, we find VOCl$_2$ to be a unique multiferroic monolayer, where tensile strain along the polar axis provides (a) high electric polarization, (b) increased ferroelectric stability and (c) enhanced insulating nature; while along the non-polar axis it causes a magnetic phase transition. Moreover, as the ferroelectricity and magnetism can be tuned independently via tensile strain along different in plane crystallographic directions, this material offers a unique opportunity to design 2D type-I multiferroic based nanoelectronic devices, having the flexibility to increases or decrease electric polarization without affecting the monolayer's magnetic properties, and vice-versa. We also find the tensile strain to be beneficial for the purpose of increasing ferroelectric and ferromagnetic Curie temperature of VOCl$_2$ monolayer. 
Our work thus reveals the versatility of VOCl$_2$, attainable via strain engineering; and this is expected to encourage further theoretical and experimental study of the VOX$_2$ family of monolayers to realize new multiferroic materials for low-dimensional technologies.

Note added: Recently, we became aware of recent theoretical works on the VOI$_2$ monolayer,\cite{Xu2020,ding2020} which predict a non-collinear magnetic ground state. However, strength of spin-orbit interactions (being proportional to Z$^4$, Z is the atomic number of the halogen atom), which is possibly responsible for the non-collinear magnetic ground state in VOI$_2$, are much smaller in case of VOCl$_2$ monolayer. Also, a recent study on VOF$_2$ monolayer,\cite{D0CP04208K} a new addition to the VOX$_2$ family, has explored collinear vs. non-collinear magnetic states, and found the former as the actual ground state. Since, Z for Cl ion (Z=17) is closer to that of F ion (Z=9), than compared to that of I ion (Z=53), we expect the VOCl$_2$ monolayer to have a collinear magnetic ground state, as reported in previous works as well.\cite{Ai2019,Tan2019}

\subsection*{Acknowledgements}
We acknowledge funding from SERB (EMR/2017/004970 and CRG/2018/002440). We also thank computer center IIT Kanpur for providing HPC facility. 
\bibliography{ref}
\end{document}